\documentclass[draft]{agujournal2019}
\usepackage{url} %this package should fix any errors with URLs in refs.
\usepackage{lineno}
\usepackage[inline]{trackchanges} %for better track changes. new option will compile document with changes incorporated.
\usepackage{soul}
\usepackage{comment}
\usepackage{xr}

\makeatletter
\newcommand*{\addFileDependency}[1]{% argument=file name and extension
  \typeout{(#1)}
  \@addtofilelist{#1}
  \IfFileExists{#1}{}{\typeout{No file #1.}}
}
\makeatother

\newcommand*{\myexternaldocument}[1]{%
    \externaldocument{#1}%
    \addFileDependency{#1.tex}%
    \addFileDependency{#1.aux}%
}
\myexternaldocument{SupportingInformationBySection}
\draftfalse

\journalname{Geophysical Research Letters}

\begin{document}
%\title{Future Changes in Mid-latitudes CAPE (Convective Available Potential Energy) and their Relationship with Moist Static Energy}

\title{Robust Relationship Between Mid-latitudes CAPE and Moist Static Energy in Present and Future Simulations \textcolor{red}{Submitted for publication in Geophysical Research Letters} }

\authors{Ziwei Wang\affil{1,2}, Elisabeth J.\ Moyer\affil{1,2}}
\affiliation{1}{Department of the Geophysical Sciences, University of Chicago, Chicago, Illinois}
\affiliation{2}{Center for Robust Decision-making on Climate and Energy Policy (RDCEP), University of Chicago, Chicago, Illinois}

\correspondingauthor{Elisabeth Moyer}{moyer@uchicago.edu}

\begin{abstract}

Convective available potential energy (CAPE), a metric associated with severe weather, is expected to increase with warming. Under the most widely-accepted theory, developed for strongly convective regimes, mean CAPE should rise following the Clausius--Clapeyron (C--C) relationship at 6--7\%/K. We show here that although the magnitude of CAPE change in high-resolution model output is only slightly underestimated with simple theories, it is insufficient to describe the distributional changes, which has a down-sloping structure and is crucial for impact assessment. A more appropriate framework for understanding CAPE changes uses the tight correlation between CAPE and moist static energy (MSE) surplus. %, the difference between surface MSE and mid-tropospheric saturation MSE. 
Atmospheric profiles develop appreciable CAPE only when MSE surplus becomes positive; beyond this point, CAPE increases as $\sim$25\% of the rise in MSE surplus. Because this relationship is robust across climate states, changes in future CAPE distributions can be well-captured by a simple scaling of present-day data using only three parameters.

\end{abstract}

\section{Introduction}

% PARA 1
% EJM notes Importance of CAPE for severe weather. CAPE is integrated buoyancy, integrated difference of temperature profiles of adiabatically rising parcel and background environment; can get explosive convection in moist conditions if the background atmosphere is very far from moist adiabat. Potential changes therefore important for socioeconomic impacts.
Convective Available Potential Energy (CAPE), loosely defined as the vertically integrated buoyancy of a near-surface air parcel, %\ziwei{some papers used the term ``loosely defined'' if we don't include the levels to integrate}
is a metric closely associated with the extreme convective weather events that can cause substantial socioeconomic damages \cite<e.g.,>{johns_severe_1992}. CAPE is derived from the difference between the temperature profile of a parcel rising pseudo-adiabatically from the surface and that of the background environment \cite{moncrieff_dynamics_1976}, which determines the maximum possible updraft velocity during undiluted ascent. 
%While CAPE can be defined with respect to any parcel initiation level, surface-based CAPE is the most commonly adopted definition for meteorological metrics and future projections (e.g., \citeA{brooks_spatial_2003,trapp_changes_2007,chen_changes_2019}). CAPE can accumulate to a large magnitude and produce explosive convection whenever background conditions deviate substantially from the moist adiabat, which is possible only in warm, humid surface conditions where the moist and dry adiabats differ strongly.
In meteorology, CAPE is used to predict thunderstorm events and in particular hail \cite{groenemeijer_sounding-derived_2007,kunz_skill_2007,kaltenbock_evaluation_2009}. Studies have also used the covariate of CAPE and wind shear to explain differences in thunderstorm frequency across locations  \cite{brooks_spatial_2003,brooks_climatological_2007} or across climate states \cite{trapp_transient_2009,diffenbaugh_robust_2013}.

%\liz{PARA 5 - validation in models}
% Discuss model studies, whether they back up Romps or not, given numbers- most are a bit above C-C. Also, are they considered CAPE or only in convective regimes? This might be several paras.
Studies of CAPE in observations have tended to focus on decadal-scale trends, often finding large increases. For example, \citeA{gettelman_multidecadal_2002} found trends equivalent to $\sim$50\%/K in 15 tropical radiosonde stations. (See SI Section S1 %\ref{section:obs_studies}
for a wider review.) Model studies of CAPE under climate change have tended to produce smaller effects. Several recent studies that simulate the tropics using convection-permitting models (0.2--4 km resolution) without advection, i.e.\ approximating radiative-convective equilibrium, find CAPE increases of 8\%/K \cite{muller_intensification_2011}, 8\%/K \cite{romps_response_2011}, 12\%/K \cite{singh_influence_2013}, 7\%/K \cite{seeley_tropical_2015}, and 6--7\%/K from theory \cite{romps_clausius_2016}. 
%with a temperature increase produced by CO$_2$ forcing.
Analyses of coarser-resolution global models have found even smaller changes in the tropical W.\ Pacific, of %$\sim$5\%/K
$\sim$4.5\%/K \cite[at 4$^\circ$ x 5$^\circ$]{ye_cape_1998} and $\sim$5\%/K %$\sim$5.5\%/K 
\cite[at 1$^\circ$]{chen_changes_2019}. In the mid-latitudes, 
%where advection plays a substantial role, 
changes may be larger: \citeA{chen_changes_2019} show $\sim$10\%/K %$\sim$12\%/K 
over a selected region of the continental United States.

% THEORY 1
Theoretical frameworks to explain climatological CAPE %changes %XXX I removed this
fall into two groups. One approach assumes that environmental profiles are fully determined by surface temperature, 
and predicts the background environmental temperature profile by considering the effects of convective entrainment.
%using the assumption that entrainment makes the actual in-cloud buoyancy in an ascending convective plume much smaller than CAPE \cite{singh_influence_2013, romps_clausius_2016}. 
% PARA 4 theories and gap
% Can't find a precursor paper that talks about CAPE change theory before SO13. There are papers that propose theories for CAPE, including EB96 and Rene and Ingersoll 1996, and Mapes 2000, both are rejected by Romps 2015. 
% EJM - is my lead-in sentence valid? - ZW: This sounds that we are saying previous work does not involve dependence on temperature. 
% -- INTRO TO OLDER PARA --Recent work on theoretical CAPE changes has focused on predicting the background environmental temperature profile by considering the effects of convective entrainment.
\citeA{singh_influence_2013} proposed a ``zero-buoyancy model'' based on the assumption that entrainment makes actual in-cloud buoyancy in an ascending convective plume small relative to CAPE, and \citeA{singh_increases_2015} evaluated its applicability in radiative-convective equilibrium simulations.
\citeA{zhou_conceptual_2019} extended the model to use an ensemble of plumes. The zero-buoyancy concept is intended to represent convective regions such as the tropics, where environmental temperature profiles are largely set by convection, with horizontal advection playing a negligible role. % \citeA{singh_increasing_2017} confirms the importance of entrainment with GCMs of different levels of complexity, showing the value of simple model applied to strongly precipitating regions only between 36 N/S. 
%\citeA{singh_increasing_2017} applied the framework up to 36 N/S but in strongly precipitating regions only, and for daily mean values.
%that the changes in convective entrainment rate and tropospheric lapse rate are important for changes in extreme CAPE.} \citeA{romps_clausius_2016} proposed an analytical solution to this framework and compared it with the numerical solution under radiative-convective equilibrium (RCE); both solutions suggest that CAPE should increase by approximately Clausius--Clapeyron (C--C) scaling ($\sim$6--7\%/K) across a wide range of surface temperatures. %This prediction is generally taken to apply to both short-term temporal variations and long-term forced climate changes.
It would not be expected to explain variations in CAPE across space or on short timescales over mid-latitudes land.%, which are known to be affected by advection and local diabatic processes. %However, zero-buoyancy theories would not be expected to explain the small-scale temporal and spatial variations in CAPE over mid-latitudes land. 

A second approach, which may be more generally applicable, % appropriate for mid-latitudes, 
treats surface and mid-tropospheric conditions as independent variables. % and explains CAPE by assuming an effective efficiency of conversion of convective heat flux to potential kinetic energy
%\cite{emanuel_moist_1996, agard_clausiusclapeyron_2017, li_midlatitude_2021}. %Surface humidity is also considered independent and not prescribed by temperature. 
Early efforts sought to characterize empirical relationships in CAPE as a function of near-surface temperature and moisture \cite{williams_analysis_1993, ye_cape_1998}. %described a linear relationship of observed CAPE in radiosondes with wet-bulb potential temperature ($\Theta_w$). 
\citeA{emanuel_moist_1996} (henceforth EB96) considered the moist static energy $h$ instead and described the relationship as 
\begin{equation}
    {\mathrm CAPE} \ = \ A \cdot (h_s - h_m)
\end{equation}
where $h_s$ and $h_m$ are moist static energy (MSE) at near-surface (boundary layer) and mid-troposphere, respectively. %their difference is termed the ``MSE deficit''. 
The dimensionless constant $A$ in EB96 reduces to $(1 - \overline{T}/T_s)$, analogous to a Carnot efficiency, where $T_s$ is the near-surface temperature and $\overline{T}$ relates to the temperature of those levels emitting radiation to space. In this perspective, CAPE represents the maximum possible kinetic energy that could be generated given a heat transfer of $(h_s - h_m)$. 

Recent work has further extended on EB96 and tested applicability to mid-latitudes CAPE. \citeA{agard_clausiusclapeyron_2017} (henceforth AE17) and \citeA{li_midlatitude_2021} use a similar functional form but slightly different formulations for the slope $A$ and for the `threshold' term. %$h_m$ in EB96 is replaced by the mean free tropospheric dry static energy. %that the environmental lapse rate is important in setting the slope $A$ of local profiles during high CAPE episodes.  
%use a similar framework but assumes the surface and the free troposphere are largely decoupled when transient CAPE peaks happen in the mid-latitudes, and use free tropospheric dry static energy instead of $h_m$ in Equation 1 as the reference threshold for CAPE development. \citeA{li_midlatitude_2021} further express the constant $A$ as $\frac{\Gamma_d}{\Gamma_{ft}} ln( T_\mathrm{LNB}/T_s)$, where $\Gamma_d$ and $\Gamma_{ft}$ are dry adiabatic and free tropospheric lapse rate, and $T_\mathrm{LNB}$ is the temperature at the level of neutral buoyancy. %In both cases $A$ is effectively the Carnot efficiency if the atmosphere is treated as a heat engine, $\sim$30\% for mid-latitudes summer. %%% People argued that it will overestimate but never tested. 
\citeA{li_midlatitude_2021} confirm that their model broadly captures both the spatial pattern and diurnal variation of CAPE in renalysis data over the continental United States. %as long as the environmental lapse rate and a scaling factors are incorporated into the slope $A$.
These theories do not fully predict future CAPE, since they provide no guidance on future changes in the threshold term relative to $h_s$, i.e.\ on changes in the shape of the environmental temperature profile. However, because all are grounded in simple mathematical definitions -- for moderately convective conditions, a linear CAPE dependence on surface MSE is a necessary consequence in any dataset where mid-tropospheric conditions are decoupled from the surface -- they should provide a useful framework for understanding model-projected changes.

% \cite{donner_three-dimensional_1999}
In this work we use a modified formulation with a different threshold term. Mathematically, CAPE is proportional to the vertically integrated difference between $h_s$ and the local ``saturation MSE'' $h^*_z$, %the MSE that midtropospheric air (at $T_z$, $p_z$) would have if it were saturated, %XXX I removed this 
neglecting the virtual temperature effect and difference in $q^*$ between parcel and environment \cite{emanuel_atmospheric_1994, randall_book_2012}. 
If we assume the shape of the environmental temperature profile does not vary strongly with $h_s$, the definition of CAPE can be reduced to:
\begin{equation}
    CAPE = A \cdot (h_s - h^*_{m})
\end{equation}
 %Since specific humidity falls strongly with temperature, $h^*_z$ quickly approaches $h_z$ with altitude. 
where $h^*_{m}$ is the minimum value of mid-tropospheric saturation MSE, and we term the difference $h_s - h^*_{m}$ the `MSE surplus'. The value of A must be determined empirically, and because its value depends on the shape of environmental profiles, it does not necessarily remain constant between climate states.

%\liz{PARA 7 State purpose of paper .... No clarity yet on projected increases in CAPE, or establishment of under what conditions C-C scaling would hold. The goal is to evaluate from the high-resolution model runs whether C-C scaling holds in future climate states in mid-latitudes, where most people live. Consider summertime North America, where precipitation is strongly convective.}
Despite the interest in understanding potential future CAPE increases, few studies have systematically evaluated these frameworks, especially in the continental mid-latitudes where severe thunderstorm impacts are greatest. %, where more people live and the extreme events do the most severe damage. 
In this work, we diagnose CAPE relationship to surface and mid-tropospheric conditions in both observation and high-resolution convection-permitting model simulations of continental North America, to determine what aspects of the relationship are robust under climate change. Our goal is to quantify projected CAPE changes in the mid-latitudes and to provide a simple framework that explains them. 
\bigskip

\section{Data Description and Methodology}
\subsection{Data Description}

The {convection-permitting} model output used here is a paired set of present and future dynamically downscaled simulations over continental North America from the Weather Research and Forecasting model (WRF, version 3.4.1) run at 4 km resolution. %XXX I edited the following two paras to shorten it (combined some sentences)
Both runs are described in \citeA{liu_continental-scale_2017}, and model output is available from NCAR Research Data Archive ds612.0 \cite{Rasmussen_Liu}. The present-day simulation (CTRL) is forced by ERA-Interim reanalysis for initial and boundary conditions; the future simulation is a pseudo-global-warming (PGW) scenario that applies a spatially varying offset to ERA-Interim based on the CMIP5 multi-model mean projection under RCP8.5. In both runs, spectral nudging is applied to levels above the planetary boundary layer. Note that hot and dry biases over the Central U.S.\ lead to a small underestimation of CAPE in the high tail by 6--10\% \cite{liu_continental-scale_2017, wang_reanalyses_2021}, but this bias does not necessarily affect fractional future changes.

In this work, we use the years 2001--2012 and equivalent future period. For `paired' comparisons we match each profile in CTRL with its equivalent in PGW. 
We calculate surfaced-based CAPE and subset to 80 grid points that match the International Global Radiosonde Archive (IGRA) weather stations as in \citeA{wang_reanalyses_2021}.  %\ziwei
See SI Section S2 %\ref{section:radiosonde} 
for spatial distribution of stations and further model validation. %, but note that the month of February 2005 in the PGW run is removed due to missing surface 2D fields. %XXX I removed this -- we don't show winter results in main text now
Most analyses here use observations in summertime (MJJA) only, when convection is most active, following \citeA{sun_evaluation_2016} and \citeA{rasmussen_changes_2017}. %We calculate CAPE from surface values, which is shown by \citeA{wang_reanalyses_2021} to be a reasonable approach for studies focusing on high CAPE. %Some supplementary figures show wintertime (NDJF) as a non-convecting case in comparison with summertime. %We have shown that using 80 stations, with a temporal coverage of 12 years provides a good sampling (472,320 profiles) to produce a CAPE distribution that is representative of the entire domain.

\subsection{Methodology} %%% Or subsetting and averaging
To maintain the focus on highly convective conditions, many  comparisons here involve values for profiles above the 73rd quantile in CAPE, which corresponds to CAPE $>$1000 J/kg in CTRL (e.g.\ Figure 3 and Figure 4, left). When computing linear fits, we use orthogonal distance regression (ODR) because it is most appropriate in conditions where errors in both dependent and independent variables matter. 
When computing fractional changes between CTRL and PGW climate states, we define them as $\ln$(PGW/CTRL)/$\Delta$T.
See SI Section S3 for details on subsetting and averaging, and \citeA{schwarzwald_changes_2021} for discussion of ODR.

\subsection{Synthetic profiles}
\label{section:synthetics}
%% Construction of surface conditions
We construct five synthetic CAPE distributions to help understand the minimal information needed to realistically reproduce future distributional changes. %: the constant offset (Const. offset), the Clausius--Clapeyron scaling (C--C scaling), the zero-buoyancy model (SO13), and the 2-param synthetic. 
All are constructed based an assumed 3.92 K surface temperature increase, the mean change for profiles above the 73rd CAPE quantile. (Note that this change is smaller than the 4.65 K average for the entire dataset; see SI Section S3.) %\ref{section:subsetting_averaging}.) 
All cases but \#4 take the CTRL profiles and CAPE values as the baseline. One case (\#1) is a simple transformation of the CTRL CAPE distribution, and three (\#2--4) require re-calculating CAPE for a set of synthetic atmospheric profiles. %nd one (\#4) directly predicts the fractional change in CAPE.
See SI Section S4 for further details.
\begin{enumerate}
\item For {\it Clausius-Clapeyron scaling}, shown for illustrative purposes only, we simply multiply each CTRL CAPE value by 1.27 ($e^{0.061 \cdot 3.92}$), where 6.1\%/K is the C--C change for the mean temperature of high-CAPE profiles, 301.8 K.  %XXX I deleted this
We omit several systematic changes that largely cancel: C--C would be changed by -0.4\%/K by including the projected reduction in surface RH, by -0.1\%/K by treating profiles separately, and by +0.6\%/K by incorporating the rise in the Level of Neutral Buoyancy (LNB). %Treating profiles separately and taking the mean C--C change would -0.1\%/K
%If each profile were treated separately, their temperature variation  would produce a range of 5.7 to 6.7\%/K, but this effect largely averages out: for the top CAPE percentile, the effect would be only -0.1\%/K.
%Including one would : the small change in surface RH in these model runs ($\sim$ -0.9\%)  
\item For the {\it constant offset} case, we add 3.92 K to each CTRL profile at each level from surface to 200 hPa, near the level of neutral buoyancy in the mean CTRL profile. From 200 hPa we linearly interpolate to zero change at 75 hPa. We also adjust surface RH by -0.9\%, the mean change above the 73rd CAPE percentile.
\item  For the {\it lapse rate adjustment synthetic} case, we modify the {\it constant offset} procedure to also include a change in lapse rate. That is, we linearly interpolate between the {3.92 K} surface warming and a similarly derived 200 hPa warming of {4.94 K}.
%The warming is held constant above 200 hPa until 75hPa, where the temperature difference is set to zero}. 
We apply the same surface RH adjustment as in {\it constant offset}.
%\item  For the {\it SO13} case, we add 3.92 K to surface temperatures and calculate environmental profiles using the zero-buoyancy model of \citeA{singh_influence_2013}. We construct profiles in both CTRL and PGW environments, so that the theory provides a self-consistent prediction of changes. 
\item  For the {\it SO13} case, we add 3.92 K to surface temperatures and calculate a climatological mean profile using the zero-buoyancy model of \citeA{singh_influence_2013}. We use an entrainment rate of 0.62 and column RH of 0.44. We construct profiles in both CTRL and PGW environments, so that the theory provides a self-consistent prediction of changes. 
\end{enumerate}

\section{Results}
\subsection{Changes in CAPE distributions}
%%% Where should the CAPE decompose - surface and environmental profile argument go? It is used in Figure1 already.
We begin our analysis by asking: in mid-latitudes model projections, how much and how does CAPE change with warming? Over the entire dataset, mean CAPE rises 61\% between CTRL and PGW, from 684 to 1103 J/kg, yielding a 10\%/K increase given the mean surface temperature rise of 4.65 K (assuming incremental changes). 
The mean change may not be the most relevant metric, however, since mid-latitude CAPE distributions are zero-inflated even in the convective summertime, and the strongest temperature changes occur in conditions where CAPE is small or zero. %(See SI Section S5.1 for present and future distributions.) %XXX I removed this
An alternate approach that emphasizes changes in higher-CAPE conditions is to take an orthogonal regression to the density distribution of paired profiles in present and future runs (Figure 1, left, solid line). This distribution shows a clear shift upwards, even though weather systems are not identical in the two runs and the scatter is therefore large. The regression slope gives a CAPE increase of 45\% or 8.0\%/K, slightly larger than Clausius Clapeyron (6.1\%/K). 
By contrast, the {\it constant offset} synthetic overpredicts CAPE increases (11.7\%/K) and the {\it SO13} theory underpredicts them (5.8\%/K); see SI Section S5.1. %When the LNB increase (0.6\%/K) is considered as suggested in \citeA{romps_clausius_2016}, the prediction from C--C scaling would better match modeled results. 

\begin{figure}[h!]
    \centering
    \includegraphics[width = \linewidth]{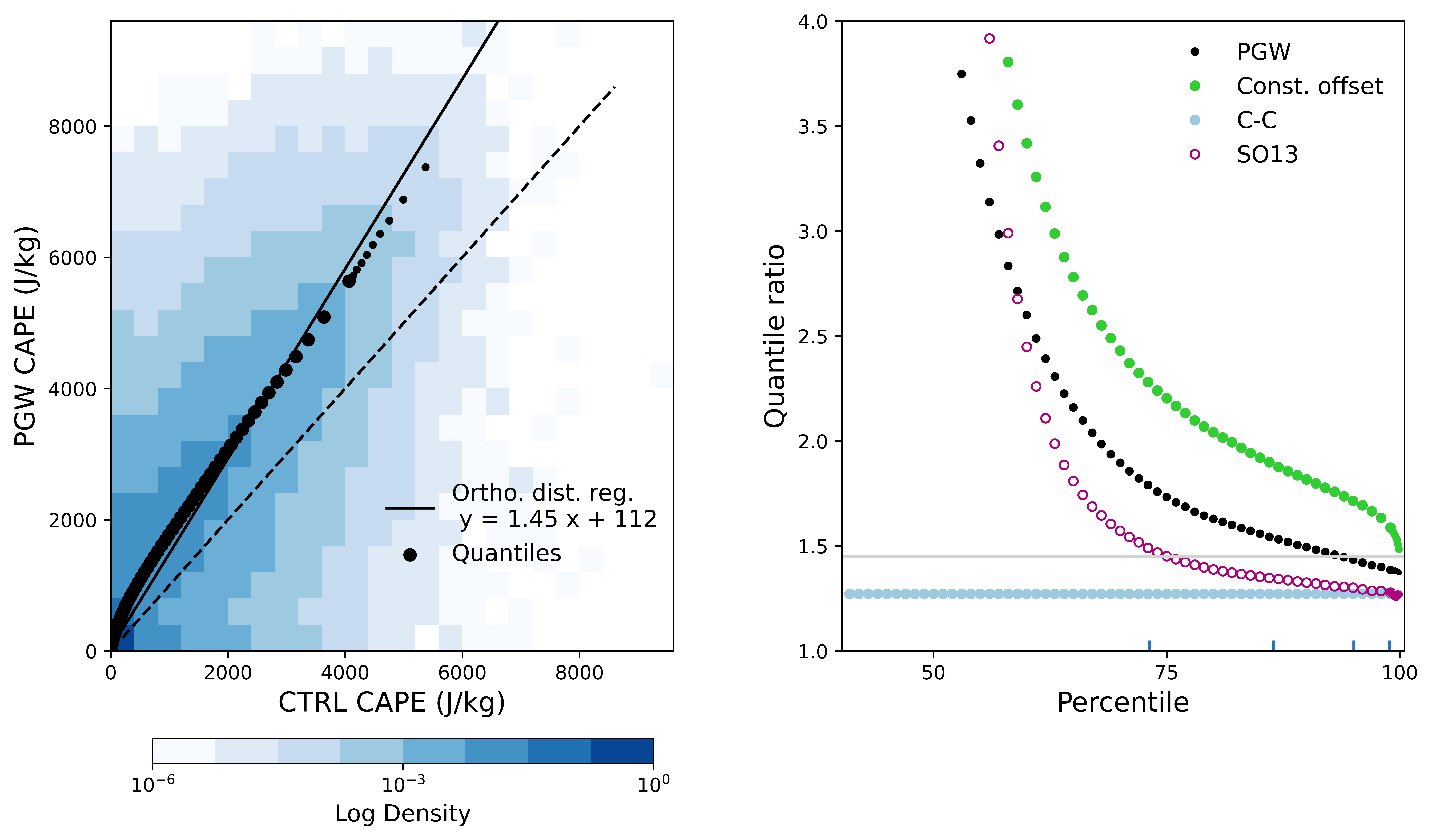}
    \caption{(Left) Comparison of CAPE in present (CTRL) and future (PGW) model runs as a density plot of paired profiles (see Methodology), using all pairs where both have nonzero CAPE. Dashed line is the one-to-one line; solid line is the orthogonal regression; and dots are quantiles of the distribution (large dots, $\Delta$ = 1\% increments from 0-0.99; small dots $\Delta$=0.1\% above 0.99). (Right) Quantile ratio plot, constructed by taking the ratio of future CAPE quantiles over those of present climate from actual model output
    %comparing future/present CAPE quantiles from actual model output
    (black, dots as in L.\ panel), and three synthetic datasets: {\it C-C scaling} (light blue), {\it constant offset} (limegreen), and {\it SO13} (purple). All data are used and zeroes are included. For internal consistency, {\it SO13} changes are computed relative to its own CTRL distribution; see methods for details. Gray horizontal line marks the mean CAPE fractional change from the orthogonal distance regression line in left panel. %Open triangles show the change in CAPE of the mean 00 UTC profile, for model output (black) and SO13 (purple). Their x-axis placement is at the quantile of the CTRL distribution that matches mean CTRL CAPE (447 J/kg). %The {\it C-C scaling}      constructed from CTRL CAPE and a $\Delta$T shift -- {\it C-C scaling} (light blue) and {\it constant offset} (limegreen) -- and for comparison,  {\it SO13} (purple).  see Methods.  The {\it C-C scaling} is intended for mean climatological shifts, but is plotted as scatters for comparison purposes. %\ziwei{Let's make sure we properly explain SO13 scatters, otherwise I still think we should leave it to supp docs.} 
    Four vertical tick bars mark the percentiles matching 1000, 2000, 3000, and 4000 J/kg (73.2\%, 86.5\%, 95.1\%, and 98.9\%, respectively). We begin the x-axis at {40}\% to omit quantiles where CTRL CAPE is zero. Model future CAPE changes resemble a constant offset with a small lapse rate adjustment.  %Pairing in the left panel suggests that there’s a lot of randomnesses produced, and synoptic patterns are not fully paired between the two runs. 
    }
    \label{fig:heatmap_qr}
\end{figure}

%We evaluate CAPE changes under warming in two ways. 
The orthogonal regression implicitly assumes that the change in CAPE distributions is a simple multiplicative shift. To test this assumption, we also show in Figure 1 a quantile regression, which compares individual quantiles of CTRL and PGW distributions. The future CAPE distribution is in fact narrower than in the simple multiplicative case. Comparing to the orthogonal regression, the lower quantiles lie above the 45\% line and the most extreme quantiles ($> \sim$3000 J/kg) below it (left panel, dots). 
%% EJM - "narrows relative" is accurate, but maybe not clear. The distribution may still widen, just less than expected given the increase. That's what the downward slope on the quantile ratio means. I don't know if the CAPE distribution actually narrows or not.
This narrowing effect is even more clear in a plot of the quantile {\it ratio} of future vs.\ present-day CAPE (right, black); it manifests as a downward slope.
Both the {\it constant offset} (green) and {\it SO13} (purple) cases also show similar narrowing, despite their different mean predicted changes.
%based on surface temperature and RH also exhibit distributional narrowing, but underestimate CAPE changes. 
%and {\it C-C scaling} cases). 
Distributional changes in model CAPE therefore resemble an offset with a small lapse rate adjustment that lowers CAPE. 

%Because the {\it SO13} theory was developed to represent the mean profile in a highly convective environment, we also show the present-future CAPE change of an averaged profile in each simulation. Because most mid-latitudes profiles are non-convective (mean CAPE of 24 J/kg in our present-day simulation), we restrict the comparison to 00 UTC (late afternoon locally; mean CAPE of 447 J/kg). The underprediction of {\it SO13} remains substantial,  at  5.1\%/K vs.\ 17.6\%/K in model output (triangles in Figure 1, right).

Because the {\it SO13} theory was developed to represent the {\it mean} profile in highly convective conditions, we also test whether it can capture the present-future CAPE change of the averaged late-afternoon (00 UTC) profile in our simulations, but the underprediction remains substantial. (See SI Section 5.1.) Changes in mid-latitudes lapse rates require a new explanatory framework. %XX not great words, just a placeholder

\subsection{Changes in environmental profiles}
% Figure 2
%Future CAPE changes are the product of a combination of an increased sampling of warmer and wetter conditions, which serve to increase CAPE, and slightly reduced lapse rates, which have the opposite effect. % ZW - I prefer a shorter version of ''the product of a combination of an ...''

\begin{figure}
    \centering
    \includegraphics[width = \linewidth]{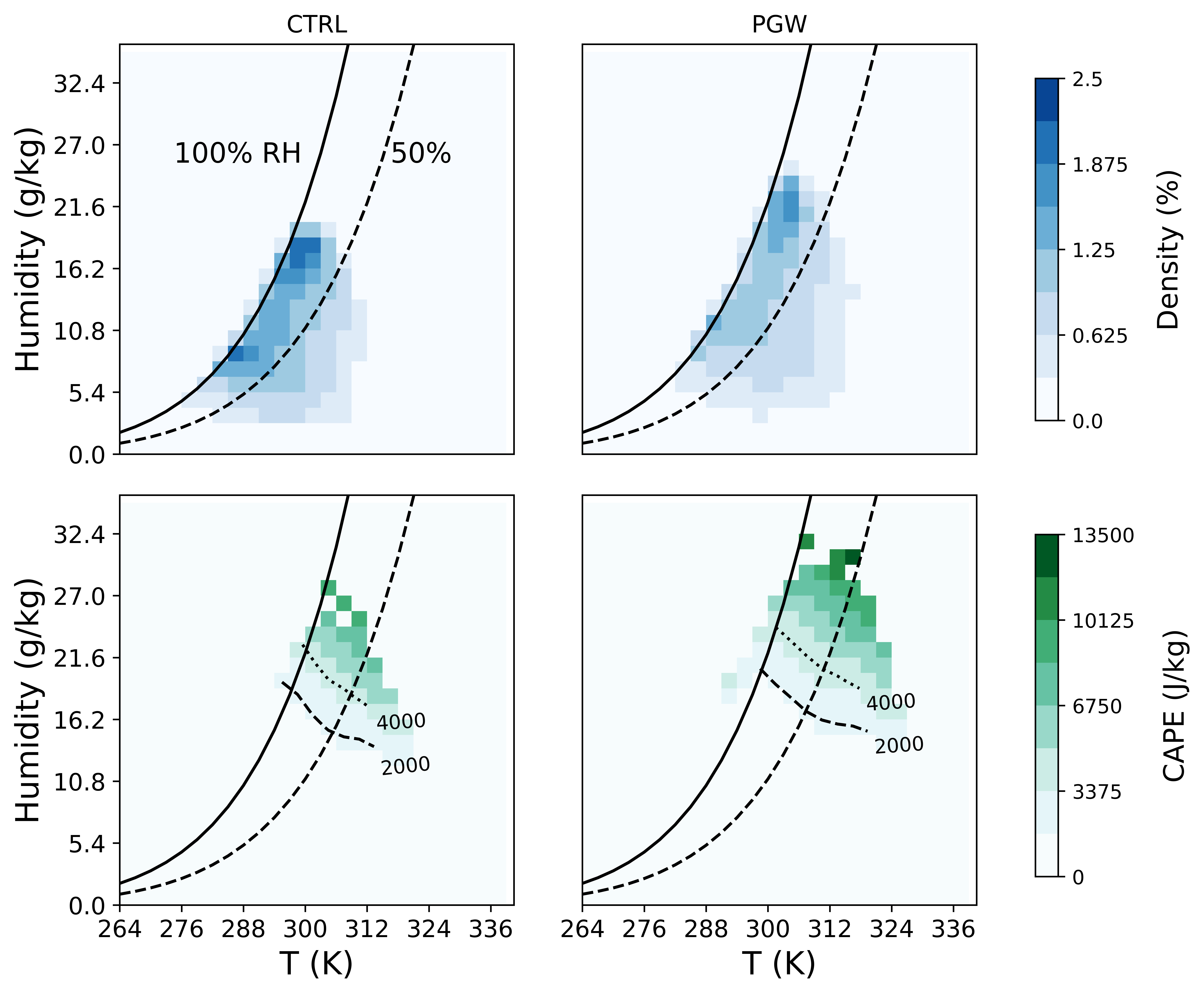}
    \caption{
    (Top) Density heatmap of T--H bins sampled and (bottom) of mean CAPE in each T--H bin, in CTRL (left) and PGW (right) runs during summer (MJJA). Bins shown are all those with 3 or more observations. Solid and dashed lines mark RH of 100 and 50\%. In bottom row, dashed/dotted lines mark CAPE contours at 2000 and 4000 J/kg (with contours cut off at RH=100\% to avoid artifacts). Both future distributions move up and to the right. The PGW run samples higher maximum temperatures (top), which in fixed environmental conditions would lead to higher CAPE by $f_{\mathrm samp}$ = 2.2, but CAPE contours also shift (bottom), reducing CAPE changes by $f_{\mathrm samp}$ = 0.62. Note that CAPE contours resemble those of moist static energy (SI Section S5.2); their future shift means that higher MSE on average is required for a given CAPE value. %See SI Section S5.2 for how well CAPE and MSE contours align onto each other. %XXX I removed this
    }
    \label{fig:heatmap_density}
\end{figure}

%\ziwei{I suggest use the word ''environmental lapse rate effect''.}
To quantify the effect of changing {environmental lapse rates} on future CAPE, we examine mean CAPE in surface temperature and humidity (T--H) space following \citeA{wang_reanalyses_2021}. Since surface T and H uniquely define the moist adiabat on which a parcel rises, a change in CAPE for a given T,H is due only to an altered environmental profile. This approach allows decomposing CAPE changes into two governing factors: $f_{\mathrm samp}$ is the fractional change that would result from only changed surface T,H sampling (Figure \ref{fig:heatmap_density}, top row) and $f_{\mathrm env}$ is that resulting from only changes in environmental profiles (Figure \ref{fig:heatmap_density}, bottom row). Both factors are defined for CTRL CAPE $>$1000 J/kg conditions.
% USE FOR CAPTION compare CAPE in profiles grouped by surface temperature and specific humidity under current and future climate. 
%% In these model runs, increased sampling of warmer surface conditions in PGW would more than double CAPE if lapse rates did not change ($f_{\mathrm samp}\sim 2.2$). However, CAPE contours shift strongly in the PGW run, so that warmer or wetter surface conditions are required to achieve the same CAPE. The average $f_{\mathrm env}$ is 0.64 for T,H conditions associated with CTRL CAPE $>$1000 J/kg, i.e.\ future environmental changes alone would damp potential CAPE increases by 36\%. 
In these model runs, increased sampling of warmer surface conditions in PGW would more than double CAPE from its CTRL value ($f_{\mathrm samp} \sim$ 2.2) if lapse rates did not change. However, CAPE contours shift strongly in the PGW run, so that warmer or wetter surface conditions are required to achieve the same CAPE. If T,H sampling remained the same, CAPE would fall by a third due to environmental effects alone ($f_{\mathrm env} \sim 0.64$). The combined effect is $f_{\mathrm samp} \cdot f_{\mathrm env} = 1.40$, close to the 1.45 derived from orthogonal regression in Figure 1. (See SI Section S5.2 for details on calculations.) 

%\ziwei{Do we want to hint that the increase is mainly due to humidity increase somewhere?}

The effects seen in Figure \ref{fig:heatmap_density} do not necessarily mean there is substantial excess warming at altitude. Most of the environmental damping of potential CAPE increases occurs even in the {\it constant offset} case of uniform warming, because present-day environmental profiles are correlated with surface temperature. Since upper tropospheric temperature is relatively homogeneous, extreme local surface temperature necessarily implies a steep lapse rate. Under climatological warming, surface temperatures that were previously extreme become associated with more normal lapse rates instead. For this reason even the {\it constant offset} case shows an $f_{\mathrm env}$ of {0.77}, i.e.\ apparent potential CAPE increases are damped by {23\%} by this covariance effect alone. (The total derived CAPE change in {\it constant offset} is {1.71}, close to its orthogonal regression slope of 1.72.)

Excess warming at altitude is therefore required only to explain the residual difference between effects in PGW ($f_{\mathrm env}$ = {0.64}) and in {\it constant offset} ($f_{\mathrm env}$ = {0.77}). Changes in temperature profiles between present and future runs are in fact very subtle. If the entire dataset is averaged, warming is actually greater at surface than at altitude ($\Delta$T$_s = 4.65$ K and $\Delta$T$_{200} = 4.05$ K), an effect that would tend to amplify CAPE. However, as discussed in Methods, when data is subdivided to include only conditions that can produce substantial CAPE, lapse rate changes are weakly positive ($\Delta$T$_s = 3.92$ K and $\Delta$T$_{200} = 4.94$). That is, in conditions favorable for convection, future environmental changes should slightly dampen the CAPE increase expected from surface warming alone.

\begin{comment}
\begin{figure}
    \centering
    \includegraphics[width = \linewidth]{Figures/Figure3_profile.png}
    \caption{Mean profiles of temperature (left) and  moist static energy (right) in CTRL (blue) and PGW (red). Solid lines show mean profiles for four subsetting thresholds based on CAPE thresholds: from lightest to darkest, all cases and $>$100, 1000, and 4000 J/kg. 
    %(See SI Figure \ref{fig:prof_tmpc_diff} for comparison to subdivision by CAPE quantiles.) 
    Color shading marks envelope of 10th--90th percentiles of all cases. (Left): Lapse rate changes are weakly positive when subset to CAPE $>$1000 J/kg or higher.
    (Right): MSE gradients in lower troposphere are strongly related to CAPE and greater in high-CAPE subsets, while mean mid-tropospheric MSE ($\overline{h_m}$) shows little CAPE dependence. These results suggest that MSE profiles are highly informative of CAPE. Future $h_s$ increases more than $h_m$ because of nonlinear contribution of moisture. See Table in SI Section S5.3 %\ref{section:si_figure3} 
    for values for $\Delta T_s$, $\Delta T_{200}$, $\overline{h_s}$, $\overline{h_m}$ of each subset, and fraction of $\Delta \overline{h_s}$ provided by latent heat. %See SI Section S5.3 for mean changes in sensible, latent enthalpy and gravitational potential of these subsets. 
    }
    \label{fig:profile}
\end{figure}
\end{comment}

\subsection{CAPE-MSE framework}
\label{section:cape_mse}

%%% need edit 
% the saturation MSE deficit may provide a simpler explanatory framework for future CAPE changes, as well as for their present-day distributions. 

\begin{figure}[h]
    \centering
    \includegraphics[width = \linewidth]{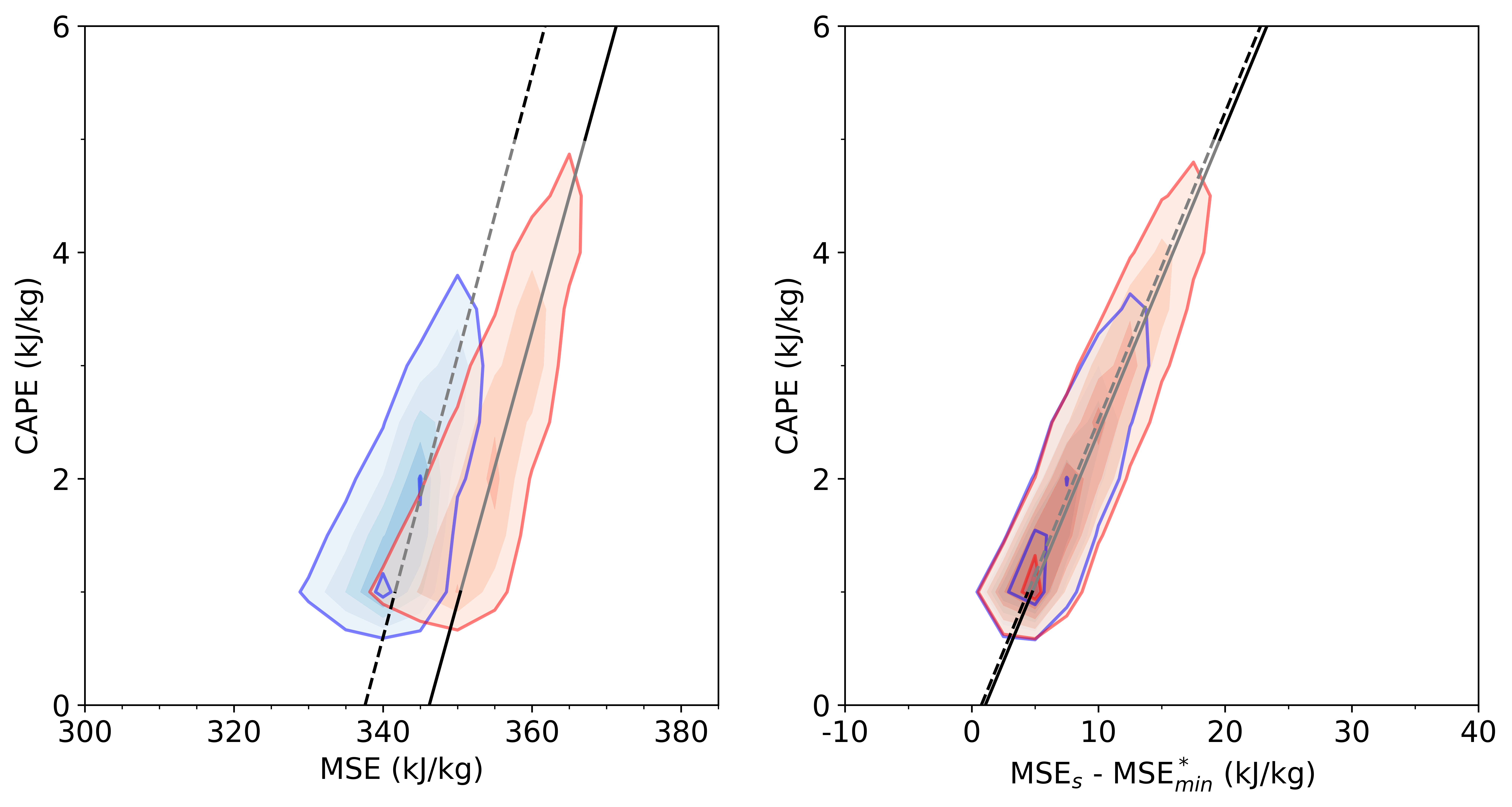}
    \caption{Relationships between CAPE in N.\ America summertime and MSE (left) and MSE surplus (right), for CTRL (blue, dotted) and PGW (red, solid) runs. Here we use all cases where CAPE is larger than 1000 J/kg. Lines are fitted orthogonal regressions. MSE surplus is calculated as $h_s$-$h^*_m$, where $h^*_m$ is the minimum saturation MSE in each profile. Color shading increments are 1.5\% for the left panel and 0.75\% for the right panel. Median in CAPE bins are used for the orthogonal regression to remove the role of uneven sampling across low to high CAPE conditions. Slopes of CAPE-MSE (left) are 0.249 and 0.239 for CTRL and PGW, respectively, and of CAPE-MSE surplus (right) are 0.271 and 0.270. }
    \label{fig:cape_mse}
\end{figure}

It is clear that CAPE in our dataset must exhibit a strong relationship with surface MSE, since the contours of CAPE in T--H space in Figure \ref{fig:heatmap_density} are closely aligned with those of MSE. (See SI Section S5.2; this effect was also shown by previous paper, e.g. \citeA{donner_three-dimensional_1999}.) %\ref{section:contours}.) %(See also SI Figure \ref{fig:cape_mse_contours}.) 
The relationship is in fact reasonably linear in each climate state (Figure \ref{fig:cape_mse}, left, which shows all CAPE values $>$1000 J/kg), but shifts as the climate warms. In both CTRL and PGW model runs, the x-intercept to the fitted regression matches the mean mid-tropospheric saturation MSE to $<0.3$\%: on average, CAPE does not develop unless surface MSE ($h_s$) exceeds saturation MSE ($h^*_m$) in the mid-troposphere. 
% save number for caption (in CTRL, 327.0 kJ/kg for the intercept vs.\ 327.7 kJ/kg for $\overline{h_m}$; in PGW, 333.3 vs.\ 334.3 kJ/kg). 
These results suggest that the more fundamental relationship is between CAPE and MSE surplus ($h_s-h^*_m$), as in Equation 2.
When CAPE is plotted against MSE surplus (Figure \ref{fig:cape_mse}, right), residual variance does indeed become smaller (24\% vs.\ 31\% for CTRL and 8\% vs.\ 26\% for PGW) and intercepts become almost zero (0.67 and 1.07 kJ/kg for CTRL and PGW, respectively).

  %, and the regression explains more than 75\% of the variances. \ziwei{Residual variance is derived by comparing the full dataset to the line regressed from binned median.}
The relationship between CAPE and MSE surplus is in fact sufficiently fundamental that it holds across climate states.  Fitted slopes are nearly identical in CTRL and PGW runs, at 0.27 (Figure \ref{fig:cape_mse}, right). In this perspective, the effects of climate change reduce to only a greater sampling of conditions with high MSE surplus. Furthermore, the relationship between CAPE and MSE surplus is robust across other temporal and spatial comparisons as well. Fitted slopes and variance explained remain similar when the dataset is divided by latitude (northern vs.\ southern stations), by daytime vs.\ nighttime profiles, by anomalously warm vs.\ cold years, or even by winter vs. summers (SI Section S5.3). Using an alternative fitting method (all samples above 1000 J/kg CAPE instead of binned median values) produces smaller slopes (0.17 and 0.16 for CTRL and PGW), but they remain consistent across all comparisons. 
%The comparison across latitudes appears analogous to that between climate states: in more southern locations, the CAPE-MSE relationship is offset since environmental temperatures are warmer, but in both southern and northern locations CAPE-MSE deficit relationships are nearly identical. 
%These relationships are also not sensitive to the exact way that $h^*_m$ is defined. Figure \ref{fig:cape_mse} uses the minimum value for each profile, but the relationship is essentially unchanged if $h_m$ is evaluated instead at a fixed pressure level of 650 hPa. \ziwei{NEED WORK}

The fact that WRF output and observations are well-described by Eq.\ 2 -- $CAPE = A \times (h_s - h^*_{m})$ -- will naturally follow if the mid-troposphere is reasonably decoupled from the surface. If variation in $h^*_{m}$ is uncorrelated with that in $h_s$, a linear relationship between CAPE and MSE surplus is a straightforward mathematical consequence. As a partial test of this condition, we plot saturation MSE profiles for data subset by a variety of CAPE thresholds (SI Section S5.4). In all conditions with any appreciable CAPE ($>$100 J/kg), the minimum of saturation MSE in the mid-troposphere remains nearly constant across subsets, suggesting that mid-tropospheric temperature and $h^*_{m}$ are not strongly coupled to surface conditions in these mid-latitudes simulations. %to produce a tight CAPE-MSE relationship in each dataset, 
%These results suggest that $h^*_m$ and $h_s$ are indeed fairly uncorrelated.

\begin{comment}
%%% earlier version: EB96, now can't compare
This concise CAPE-MSE deficit framework is consistent with the general mathematical expression in ``heat engine'' theories such as EB96. (Only one small adjustment is needed, in that the non-zero x-intercepts of the fitted relationships imply that a small MSE deficit is required before CAPE can be developed). Furthermore, the derived slopes of the CAPE-MSE deficit relationship in both model output and in radiosonde observations are consistent with that proposed in EB96. We apply the EB96 framework to the mean profile in CAPE $>$ 1000 J/kg conditions, and calculate a theoretical value of 0.18 for both CTRL model output and observations, similar to their empirical slopes of 0.17 and 0.16. (See SI Section S4.2 for details.) While the AE17 framework produces considerably larger slopes (0.32 for CTRL), the EB96 framework appears to successfully capture the ``conversion rate'' that translates MSE deficit into CAPE. 
\end{comment}

\subsection{A simple lapse rate adjustment framework}

% Figure 5
% ZW - while or "given that"? 
While theories of future CAPE based only on surface conditions do not work well in the mid-latitudes, we consider whether adding a single parameter to describe mid-tropospheric effects can yield accurate predictions of future CAPE distributions. As described in Section 2.3, we construct a transformation of present-day atmospheric profiles based on only 3 parameters: mean changes in surface temperature and humidity, and a separate value for warming at 200 hPa ($\Delta T_s$, $\Delta$ RH, $\Delta T_{200}$).
%To understand whether we can how well existing theories and simple models reproduce future changes in the CAPE distribution, 
To evaluate how well this {\it lapse rate adjustment} captures CAPE changes in actual model output, we show also results for a two-parameter transformation -- the {\it constant offset} shift with RH adjustment, which uses only mean surface $\Delta T_s$ and $\Delta$ RH -- and  %the zero-buoyancy theory of {\it SO13}, its simplification to 
for reference, a simple {\it C-C scaling} applied to each individual profile. See Section 2.3 and SI Section S5.5 for details.

\begin{figure}[h!]
    \centering
    \includegraphics[width = \linewidth]{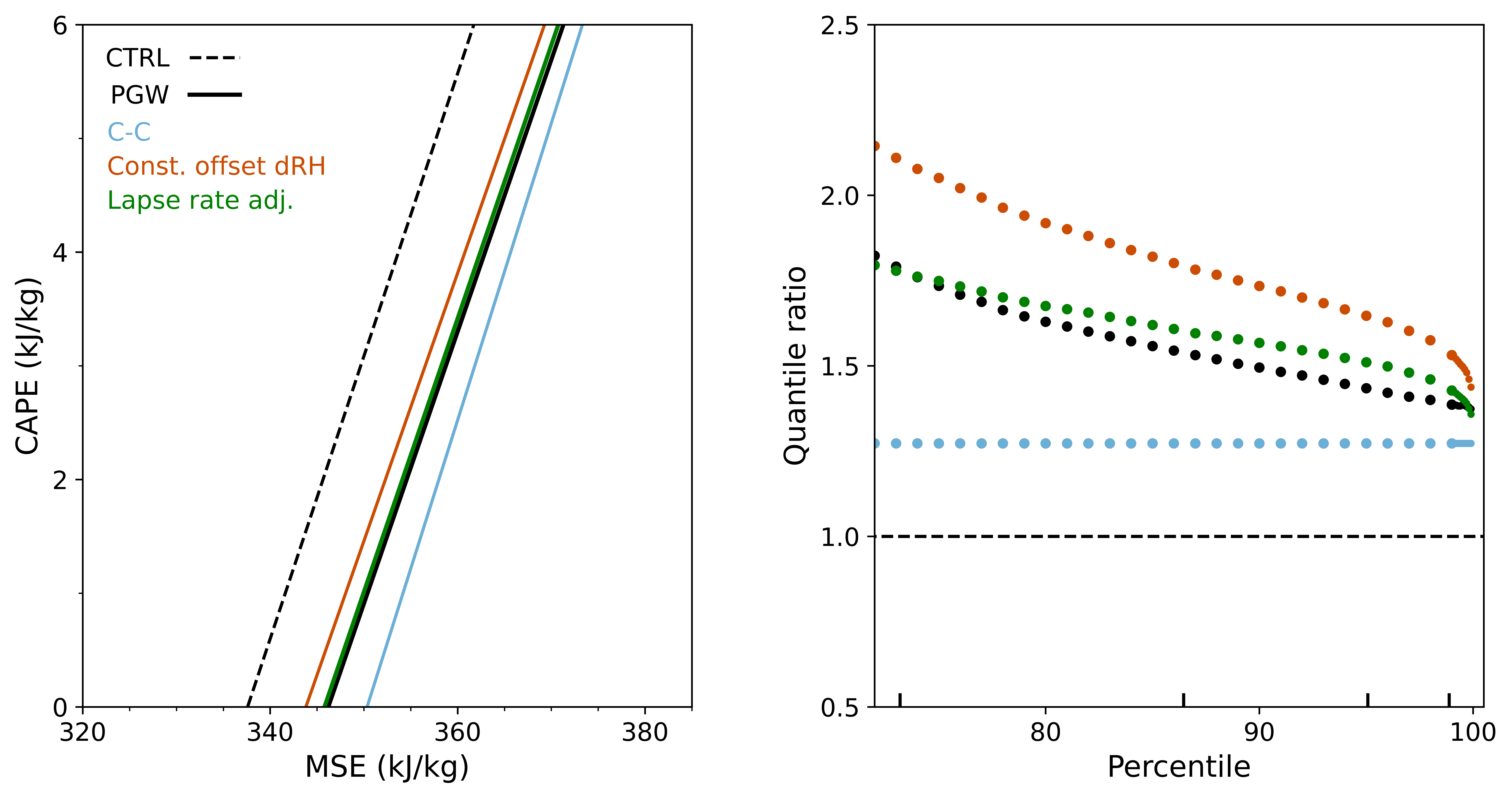}
    \caption{Comparison of present and future CAPE in model output (black) and synthetics, with those built from existing theories ({\it C--C scaling}, light blue; and from this work in the bottom row ({\it constant offset}, dark orange; {\it lapse rate adjustment}, green). (Left): Fitted regression lines of the future CAPE-MSE relationship as in Figure \ref{fig:cape_mse}. Model CTRL is shown for reference (dashed black). See SI Section S5.5 for more details, including table of slopes and x-intercepts. (Right) Future changes in CAPE as quantile ratio plots, with dots marking quantiles at 1\% increments. As in Figure 1, four x-axis ticks mark 1000--4000 J/kg, and PGW/CTRL CAPE values are on the numerator/denominator. All the synthetic future (scatters) fractional changes are referenced to CTRL. CAPE-MSE instead of the CAPE-MSE surplus framework is shown because the latter requires further assumptions about how the mid-tropospheric MSE would change. The {\it lapse rate adjustment} synthetics best reproduce future CAPE. %, with the {\it lapse rate adjustment} most closely capturing changes in the higher quantiles.
    % Slopes for CTRL, PGW, C-C, Constant offset dRH and Lapse rate adjustment are 0.249, 0.239, 0.271, 0.240, 0.236
    }
    \label{fig:synthetic}
\end{figure}

 The three-parameter {\it lapse rate adjustment} transformation does indeed capture the characteristics of future CAPE changes (in high-CAPE conditions). In the CAPE-MSE perspective (Figure \ref{fig:synthetic}, left), it realistically captures the future relationship, both in its slope and x-intercept. 
 %\ziwei{With the x-intercept representing the mid-tropospheric saturation MSE, we could deduct the preformance of CAPE-MSE surplus framework without the need of making further assumptions of mid-tropospheric changes in each synthetic.} 
 In the quantile ratio perspective (Figure \ref{fig:synthetic}, right), it reproduces both the downsloping structure and the magnitude of fractional change in the high CAPE quantiles. On a T--H diagram, {\it lapse rate adjustment} reproduces the future CAPE contours well while other transformations produce clear discrepancies (SI Section S5.5). Note that in the highest CAPE conditions, future changes in model output and in {\it lapse rate adjustment} begin to approach Clausius-Clapeyron, but remain above it. Changes in the 99th quantile are 6.9\%/K in WRF and 7.1\%/K in {\it lapse rate adjustment}, while the {\it C--C} line in Figure \ref{fig:synthetic} is shown as a constant 6.1\%/K, and would be similar even if treated more realistically. 
 (See Methodology, and SI Section 5.5 for more extensive comparisons.) While mid-latitudes CAPE is too complex to be treated with simple scalings, a relatively straightforward 3-parameter transformation appears to reproduce its full distributional change in a future warmer climate.

\newpage
\section{Discussion}

%------------ NEW -----------------------

% para 1 on magnitude of change and problematic
Increases in severe weather events, which are associated with high CAPE, are a substantial societal concern under global warming.
We show here that the projected increase in mean mid-latitudes CAPE in high-resolution model output is substantially higher than in theories developed under assumptions appropriate for the tropics, which are close to Clausius-Clapeyron (C--C). The discrepancy is smaller for the most extreme conditions, but even in the highest quantiles in this analysis, model CAPE changes are over 20\% above C--C. This difference translates to large changes in the projected occurrence of CAPE exceeding a given threshold. For example, incidences of summertime CAPE $>$2000 J/kg, a commonly-used threshold for severe weather, rise twice as much in model projections as in Clausius--Clapeyron scaling: from 13\% in CTRL to over 24\% in the future PGW projection, vs.\ to only 19\% under C--C scaling. %18.7\% to be accurate %This lower fractional change is not surprising, given that the slope under ``heat engine'' theory is assumed to be Carnot efficiency, while in reality the condensational heating does not happen at a single level.
% for the same temperature change

% P2 - simplicity para
%%% ZW - first sentence should be moved to the end of this para
%\ziwei{I edited this paragraph}
%Theoretical approaches based on RCE provides a useful tool to understand both the underlying physics of current climate and the changes of climatological mean profiles, based on  local surface values and assumptions of column RH and entrainment rate only. However, they are not designed to predict the CAPE in individual profiles of present-day mid-latitudes, or the distributional changes.
%To understand CAPE changes in individual profiles on climatological timescales, the stronger upper tropospheric warming (the lapse rate effect) in mid-latitudes model output must be noted. Because the model CAPE change is smallest in the uppermost quantiles, RCE-based theory most closely matches model projections in the highest-CAPE conditions, providing a useful guideline for CAPE changes in the extremely high quantiles. 

The midlatitudes apparently require a different framework for understanding CAPE changes than the convective tropics. Both the influence of advection and the strong surface diurnal variation means that mid-tropospheric values cannot be predicted from surface conditions. Furthermore, the wide range of surface conditions in the mid-latitudes continental U.S.\ mean that lapse rate effects vary spatially across the domain, with upper tropospheric warming strongest in the subtropics and lapse rates changes actually negative north of 33N (SI Section S7). Nevertheless, we find that future CAPE distributional changes can be well-captured by a simple synthetic transformation based only on three changes averaged over the entire domain ($\Delta T_s$,  $\Delta$RH$_s$, and either $\Delta T_{200}$ or $\Delta T_{650}$). 
% However, we show here that adding a single additional parameter to account for lapse rate changes is sufficient to capture the full distributional change in CAPE. This result is unexpected, 
% The simple three-parameter transformation is a useful supplement to the existing theories -- not as a full physical theory but a framework that helps understand full distributional changes in CAPE. 
These three parameters can be folded into a single metric of ``MSE surplus'', the difference between surface MSE and mid-tropospheric saturation MSE. In the model output described here, CAPE does exhibit a strong dependence on MSE surplus, as expected: in each climate state the relationship is a straightforward mathematical consequence. We show here that the relationship is robust even across climate states (empirical slopes of 0.27 and 0.26 in Figure \ref{fig:cape_mse}) implying that atmospheric structure does not change dramatically. 

%%% comment out for word limit
These results can be compared to prior theories based on analogies to heat engines. The slope $A$ can be thought of as the maximum conversion rate of MSE surplus to mechanical work. Similarly, theories such as EB96 treated CAPE as the maximum work possible given a flow of energy between hot and cold reservoirs, and therefore predicted a Carnot-like slope of $(1 - \overline{T}/T_s)$. This theoretical value can be derived by constructing a mean atmospheric profile (using all incidences of CAPE $>$ 1000 J/kg); our dataset yield theoretical slopes of 0.18 for both radiosonde observations and CTRL model output, similar to the 0.14 in \citeA{emanuel_moist_1996}. This value is lower than the empirical slopes of Figure 3 (right), but %XXX I edited this part and combined with the next sentennce %note that we use a different regressor from \citeA{emanuel_moist_1996}, $(h_s - h^*_m)$ vs.\ their $(h_s - h_m)$ and fit the binned median slope. Reproducing their methods on our data yields 
is nearly identical to slopes derived without fitting the binned median values: 0.18 for observations and 0.17 for CTRL. It appears that the heat engine framework does capture some physical constraint on CAPE, though MSE surplus $(h_s - h^*_m)$ is the more fundamental regressor. %XXX 
Note that CAPE represents only the {\it potential} production of kinetic energy, not the true conversion rate, which is affected by factors that reduced efficiency below Carnot \cite<e.g.>{romps_dry_2008}. 

Understanding how CAPE responds to CO$_2$-induced warming is a key scientific question with significant societal consequences. This work suggests that in the mid-latitudes, the decoupling of surface and mid-troposphere means that changes in CAPE can be larger than predicted by theories developed for the convective tropics. We find that a simple 3-parameter transformation captures not only future mean increases in midlatitudes CAPE but their full distributional shifts. It does remain an outstanding question how the present-day mapping of CAPE to convective updraft velocities and extreme convective events may alter under climate change. However, the strong and consistent dependence of CAPE on MSE surplus provides a simple but robust framework for predicting and understanding changes in CAPE distributions.

\bigskip 

\textbf{Acknowledgements}
The authors thank Dan Chavas, Tiffany Shaw, Funing Li, Zhihong Tan, and Osamu Miyawaki for constructive comments, and the National Center for Atmospheric Research (NCAR) for providing the WRF dataset. This work is supported by the Center for Robust Decision-making on Climate and Energy Policy (RDCEP), funded by the NSF Decision Making Under Uncertainty program, Award SES-1463644, and was completed in part with resources provided by the University of Chicago Research Computing Center.

\textbf{Data Availability Statement}
The 4-km WRF Convection-permitting model output is downloaded from NCAR RDA \url{https://rda.ucar.edu/datasets/ds612.0/} (\url{http://doi.org/10.5065/D6V40SXP}). The IGRA radiosonde data is downloaded from \url{https://www.ncei.noaa.gov/products/weather-balloon/integrated-global-radiosonde-archive} (\url{http://doi.org/10.7289/V5X63K0Q}).

\newpage

\bibliography{reference}

\end{document}